\documentclass{aa}
\usepackage{graphicx}
\usepackage[varg]{txfonts}
\usepackage[colorlinks=true,linkcolor=blue, citecolor=blue, filecolor=blue, urlcolor=blue]{hyperref}

\begin{document}

\title{Rugged magneto-hydrodynamic invariants in weakly collisional plasma turbulence: Two-dimensional hybrid simulation results}

\author{Petr Hellinger\inst{1,2} \and Victor Montagud-Camps\inst{3}}
\institute{Astronomical Institute of the Czech Academy of Sciences, Prague, Czechia
  \and
     Institute of Atmospheric Physics of the Czech Academy of Sciences, Prague, Czechia
  \and
     Department of Electromagnetism and Electronics, University of Murcia, Murcia, Spain
}

\date{ 14 August 2024 }

  \abstract
   {}
      {We investigated plasma turbulence in the context of solar wind. We concentrated
   on properties of ideal second-order magneto-hydrodynamic (MHD) and Hall MHD invariants.
   }
   {We studied the results of a two-dimensional hybrid simulation of decaying plasma turbulence 
   with an initial large cross helicity and a negligible magnetic helicity.
   We investigated the evolution of the combined energy and the cross, kinetic, mixed,
   and magnetic helicities. For the combined (kinetic plus magnetic) energy and the 
   cross, kinetic, and mixed
   helicities, we analysed the corresponding K\'arm\'an-Howarth-Monin (KHM) equation
  in the hybrid (kinetic proton and fluid electron) approximation.}
   {The KHM analysis shows that the combined energy decays at large scales. At intermediate scales,
  this energy cascades (from large to small scales) via the MHD non-linearity and this cascade partly continues via Hall coupling
  to sub-ion scales. The cascading combined energy is transferred (dissipated) to the internal
   energy at small scales via the resistive dissipation and the pressure-strain effect.
  The Hall term couples the cross helicity with the kinetic one, suggesting that the coupled
  invariant, referred to  here as the mixed helicity, is a relevant turbulence quantity. However, when analysed using 
  the KHM equations, the kinetic and mixed helicities exhibit very dissimilar behaviours to that of the combined energy. 
  On the other hand, the cross helicity, in analogy to the energy, decays at large scales, cascades 
  from large to small scales via the MHD+Hall non-linearity, and is dissipated at small scales via 
  the resistive dissipation and  the cross-helicity equivalent of the pressure-strain effect. 
  In contrast to the combined energy, the Hall term is important for the cross helicity over a wide range of scales 
  (even well above ion scales).
  In contrast, the magnetic helicity is scantily generated through the resistive term
  and does not exhibit any cascade.
}
   {}


   \authorrunning{Hellinger \& Montagud-Camps}
   \titlerunning{Rugged MHD invariants in plasma turbulence}

\maketitle
\nolinenumbers

\section{Introduction}

Fluids in a turbulent regime transfer the energy of fluctuations across scales due to their non-linear
interactions, in a process known as energy cascade.  Non-linearities not only
act upon the energy
of fluctuations, but also on other turbulence properties as well. This enables the formation of multiple
cascades in the system, with the development of one affecting that of another \citep{albi18}.
Properties of incompressible magneto-hydrodynamic (MHD) fluid are
strongly determined by three second-order (rugged) invariants:
combined (kinetic plus magnetic) energy, and cross and magnetic helicities.
While the magnetic helicity remains an ideal invariant
even in the compressible Hall MHD, the presence of the Hall term couples
the cross and kinetic helicities; a combination
of the two, the mixed helicity \cite[also called X helicity; see][]{poyo22} is the 
ideal invariant appropriate for the incompressible Hall MHD.
These ideal invariants are particularly relevant in turbulence.
Cascade and dissipation of the combined energy 
is one of the central problems of (Hall) MHD turbulence 
and those processes are influenced by the different helicities,
 which in turn may exhibit analogous cascade and dissipation phenomena 
to the energy ones. All these processes constrain and change the dynamics of turbulent systems. For
instance, cross helicity may reduce the energy cascade \citep{dobral80} and plays an important role in the generation of magnetic fields through
turbulence dynamo processes \citep{yoko99}; it may also affect the repartition of heating
between ions and electrons \citep{squial23}.

The solar wind exhibits important fluctuations of the magnetic field, plasma bulk velocity,
and other quantities over a wide range of scales \citep{kiyaal15,alexal21}. These fluctuations often
have power-law-like spectral properties and are non-linearly coupled; the solar wind system is 
an example of a turbulent system in weakly collisional plasmas \citep{brca13}. 
While slow solar wind streams have a weak average cross helicity \cite[however see e.g.][]{dabr15,shial21},
fast solar wind streams have typically relatively strong average cross helicities. 
These high-cross helicity streams are also called Alfv\'enic streams as they
exhibit important correlations or anti-correlations between the magnetic
field and the plasma bulk velocities, which are characteristic of Alfv\'en waves \citep{beda71,grapal81}.
Plasma turbulence in high-cross helicity streams
and the behaviour of the combined energy and cross helicity
 are often studied in the incompressible MHD approximation and
the energy and helicity is replaced by pseudo-energies 
in the terms of the Els\"asser variables \cite[combinations of
the magnetic field in the Alf\'en units  and the plasma bulk velocity; see][]{tual89}.
In situ observations in the solar wind, as well as numerical simulations,
indicate that the two pseudo-energies cascade in parallel, which means
that the combined energy and cross helicity cascade in parallel.
However, at variance with predictions of incompressible MHD \citep{dobral80},
the combined energy in the solar wind decreases more slowly with the radial distance than 
the cross helicity, and consequently the relative
cross helicity decreases.
It is therefore important to 
determine the influence
of compressive and non-ideal (Hall and kinetic) effects.

Compressive effects break the invariance of the combined energy
and the mixed helicity, but
the solar wind exhibits only weak density fluctuations \citep{shial21}, meaning that
 compressibility effects are likely not dominant in solar wind turbulence;
the compressible MHD turbulence simulations of \cite{montal22} behave qualitatively similar
to incompressible simulations, again in contrast with observations.
For weakly collisional solar-wind
plasmas, it is also necessary to investigate the consequences of formation of non-Maxwellian particle distribution functions,
which requires the tensor description of the particle pressure \citep{delsal16}.
One of these consequences is the pressure-strain coupling, which appears to be very important, as
it may work as an effective dissipation rate for the combined
energy \citep{yangal17,mattal20}.
In this paper, we address the question how the pressure-strain effect
influences the different helicities in order to complete our understanding
of the rugged invariants in weakly collisional plasma turbulence.   
We analysed results of the two-dimensional pseudo-spectral hybrid simulation of
 decaying, high-cross helicity plasma turbulence.
We used the K\'arm\'an-Howarth-Monin  (KHM) equation \citep{kaho38,popo98a,popo98b} in the hybrid approximation
for the energy \citep{hellal24} and we derived the equivalent equations
for the helicities.
The paper is organised as follows:
In Section~\ref{code} we describe the pseudo-spectral hybrid code and in section~\ref{eqcons}
we summarise the governing equations and their conservation properties.
Section~\ref{simul} presents the overall simulation results and
in section~\ref{KHMs} we analyse the KHM properties of
the energy (subsect.~\ref{KHM1}) and the cross, kinetic, and mixed
helicities (subsect.~\ref{KHM2}).
Finally, in section~\ref{discuss}
we summarise and discuss the simulation results. In the Appendix we complement our KHM results with 
spatial and spectral filtering techniques.

\section{Numerical code}
\label{code}
For the numerical simulation, we used the 2D pseudo-spectral version of the hybrid code
based on the model of \cite{matt94}. In this code, the electrons are
considered as a massless charge-neutralising fluid, with a constant temperature, whereas ions are
treated as particles. Ions have positions and velocities separated by  
half time steps and are advanced by the Boris scheme, which requires
the fields to be known at a half time step ahead of the
particle velocities. This is obtained by advancing the current density to
this time (with only one computational pass
through the particle data at each time step). 
The same grid is used for all the fields and their spatial derivatives,
needed for the time advance \cite[cf.][]{valeal07}, are calculated
with the fast Fourier transform \citep{frjo05}.
The particle contribution to the current density
at the relevant nodes is evaluated with bilinear weighting
followed by smoothing over three points. No smoothing is performed on
the electromagnetic fields. A small resistivity is used in the Ohm's law
to avoid an accumulation of energy at small scales.
The magnetic field is advanced in time with a modified midpoint method,
which allows time substepping to advance the field.

The units and parameters of the simulation are as follows: units of space
and time are the ion inertial length $d_i=c/\omega_{pi}$ and 
the (inverse of the) ion gyro-frequency $\Omega_i$, respectively,
 where $\omega_{pi} = ({n_p e^2}/{m_p\epsilon_0})^{1/2}$ is the
proton plasma frequency.  In these expressions, $n_p$ and $B_0$ are
 the density of the plasma protons
and magnitude of the initial magnetic field, respectively,
 while $e$ and $m_p$ are the
proton electric charge and mass, respectively; and, finally
$c$, $\epsilon_0$, and $\mu_0$ are the speed of light and the
dielectric and magnetic permeabilities of vacuum, respectively.
 The spatial resolution is $\Delta x=\Delta y=d_i/8$. There are $16,384$ 
particles per cell  for protons; $\beta_i=\beta_e=0.5$. The simulation box
is in the $x$-$y$ plane  and is
assumed to be periodic in both dimensions. The fields and moments are
defined on a 2D grid with dimensions $2048\times 2048$. The time
step for the particle advance is $dt=0.01\Omega_i^{-1}$ while the magnetic
field $\boldsymbol{B}$ is advanced with a smaller time step, $dt_B = dt/20$.
(The vector potential $\boldsymbol{A}$ is initialised from the magnetic field
assuming the Coulomb gauge and is then evolved in time in the code.)
The background magnetic field $\boldsymbol{B}_0$ is perpendicular to the
simulation plane. Following \cite{franal15b},
we initialised the system with an isotropic 2D spectrum of modes with random phases, linear Alfv\'en
polarisation ($\delta \boldsymbol{B} \perp \boldsymbol{B}_0$) over large scales  $k\le 0.1 d_i^{-1}$
with a flat 1D spectrum. The system initially has the (rms) amplitude of magnetic
field fluctuations $\delta B/B_0=0.3$, and the relative cross helicity $\sigma_c=0.6$. 
The magnetic and kinetic helicities are initially almost zero.

\section{Governing equations}
\label{eqcons}

We investigated a system governed by the following
equations for the plasma density $\rho$,
the plasma mean velocity $\boldsymbol{u}$, and the magnetic field
$\boldsymbol{B}$:
\begin{align}
\frac{\partial \rho}{\partial t}+(\boldsymbol{u}\cdot\boldsymbol{\nabla})\rho&=
-\rho \boldsymbol{\nabla}\cdot \boldsymbol{u}, \label{density} \\
\frac{\partial\boldsymbol{u}}{\partial t}+(\boldsymbol{u}\cdot\boldsymbol{\nabla})\boldsymbol{u}&=
\frac{\boldsymbol{J}\times\boldsymbol{B}}{\rho}
 -\frac{\boldsymbol{\nabla}\cdot \mathbf{P}}{\rho}
\label{HallMHDv} \\
\frac{\partial\boldsymbol{B}}{\partial t}&=
\boldsymbol{\nabla}\times \left[(\boldsymbol{u}-\boldsymbol{j})\times\boldsymbol{B}\right]
+\eta\nabla^2\boldsymbol{B}.
\label{HallMHD}
\end{align}
Here $\mathbf{P}$ is the plasma pressure tensor, $\eta$ is the electric resistivity, 
$\boldsymbol{J}$ is the electric current density, and $\boldsymbol{j}$
is the electric current density in velocity units, $\boldsymbol{j}= \boldsymbol{J}/\rho_c=\boldsymbol{u}-\boldsymbol{u}_e$ ($\rho_c$ and $\boldsymbol{u}_e$ being the charge density
and the electron velocity, respectively). We assume SI units except for the magnetic permeability $\mu_0$, which is set to one
(SI results can be obtained by the rescaling $\boldsymbol{B} \rightarrow \boldsymbol{B} \mu_0^{-1/2}$).
Equation~(\ref{HallMHD}) (except the resistive term) can be derived taking moments of the Vlasov equation for protons and electrons,
and assuming massless electrons. We added a resistive dissipation, as used in the hybrid code
\cite[cf.][]{hellal24}.

We investigated properties of the different energies and helicities
in electron--proton plasma in the hybrid approximation.
For the sum of the (proton) kinetic ($E_\text{kin}=\langle\rho|\boldsymbol{u}|^2 \rangle/2$) 
and the magnetic ($E_\text{mag}=\langle|\boldsymbol{B}|^2\rangle/2$) energies 
averaged over a closed volume (denoted by angle brackets $\langle \bullet \rangle$) 
we obtained the following budget equation:
\begin{equation}
 \partial_t \left(E_\text{kin} + E_\text{mag}\right) = -Q,
 \label{dtEn}
\end{equation}
where 
$Q=Q_\eta+\psi$ denotes the total (effective) dissipation rate consisting of 
the pressure-strain rate
$\psi=-\langle \mathbf{P} : \boldsymbol{\nabla} \boldsymbol{u} \rangle$,
 and the resistive dissipation rate 
 $Q_\eta=\eta \langle |\boldsymbol{J}|^2 \rangle$.
Following \cite{kior90}, we defined a density-weighted velocity field $\boldsymbol{w}=\rho^{1/2}\boldsymbol{u}$
and we represented the kinetic energy as a second-order positively definite quantity,
$E_\text{kin}=\langle|\boldsymbol{w}|^2 \rangle/2$.

For the averaged cross helicity
($H_c=\langle \boldsymbol{u}\cdot \boldsymbol{B} \rangle$),
we obtained the following equation: 
\begin{equation}
\partial_t H_c  =
- \left\langle \boldsymbol{\omega}\cdot\left(\boldsymbol{j}\times\boldsymbol{B}\right)\right\rangle
- \psi_{Hc} -\epsilon_{\eta Hc}
\label{dtHc}
,\end{equation}
where $\boldsymbol{\omega}=\boldsymbol{\nabla} \times \boldsymbol{u}$ is
the vorticity field,
$\psi_{Hc}=-\langle \mathbf{P}  :  \boldsymbol{\nabla} (\boldsymbol{B}/\rho)\rangle$
is the cross helicity equivalent of the pressure-strain rate 
\cite[generalisation of the pressure--dilation coupling; see][]{montal22},
and 
$\epsilon_{\eta Hc}=\eta\langle\boldsymbol{\omega}\cdot\boldsymbol{J}\rangle$
is the cross helicity resistive dissipation rate.
The first term on the right-hand side of Eq.~(\ref{dtHc}) couples the cross helicity
to the kinetic helicity.

For the (rescaled) kinetic helicity
($H_k=m\langle \boldsymbol{u}\cdot \boldsymbol{\omega}  \rangle/(2e)$),
we derived the dynamic equation
\begin{equation}
\partial_t H_k =
\left\langle \boldsymbol{\omega}\cdot\left(\boldsymbol{j}\times\boldsymbol{B}\right)\right\rangle
-\psi_{Hk},
\end{equation}
where 
$\psi_{Hk}=-m\langle \mathbf{P}  :\boldsymbol{\nabla} (\boldsymbol{\omega}/{\rho})\rangle/e$
is the kinetic helicity equivalent of the pressure-strain rate.
In the above formulas, we chose to represent the kinetic helicity with the renormalisation factor 
$m/e$, that is, the proton mass-to-charge ratio, in order to have the cross and kinetic helicities in the same units.
The mixed helicity, the sum of the cross and  (renormalised) kinetic helicities,
$H_x= H_c + H_k$, is then an ideal invariant of the hybrid system, and behaves as
\begin{equation}
\partial_t H_x =
- \psi_{Hc} -\epsilon_{\eta Hc}
-\psi_{Hk}.
\end{equation}

The magnetic helicity is a separate ideal invariant of Hall MHD as well as of the hybrid system.
In this paper, we investigate properties of a simulated system with 
periodic boundary conditions and a background uniform magnetic field $\boldsymbol{B}_0=\langle\boldsymbol{B}\rangle$.
In this case, the vector potential $\boldsymbol{A}_0$ generating
$\boldsymbol{B}_0=\boldsymbol{\nabla}\times \boldsymbol{A}_0$ is
not periodic and affects the magnetic helicity conservation 
\citep{mago82}; we studied properties 
of the modified magnetic helicity 
$H_m=\langle \boldsymbol{A}_1 \cdot (\boldsymbol{B}_1+2\boldsymbol{B}_0)  \rangle$,
where $\boldsymbol{B}_1$ and $\boldsymbol{A}_1$ are the fluctuating 
components of the magnetic and vector potential fields, 
$\boldsymbol{B}_1=\boldsymbol{B}-\boldsymbol{B}_0$
and
$\boldsymbol{A}_1=\boldsymbol{A}-\boldsymbol{A}_0$,
respectively.
For the averaged modified magnetic helicity, we obtained the following conservation properties:
\begin{equation}
\partial_t H_m = - \epsilon_{Hm} 
,\end{equation}
where 
$\epsilon_{Hm}= 2 \eta \langle\boldsymbol{B}\cdot\boldsymbol{J}\rangle$ 
is the resistive dissipation rate.

\section{Simulation results}
\label{simul}

\begin{figure}
\centerline{\includegraphics[width=8.2cm]{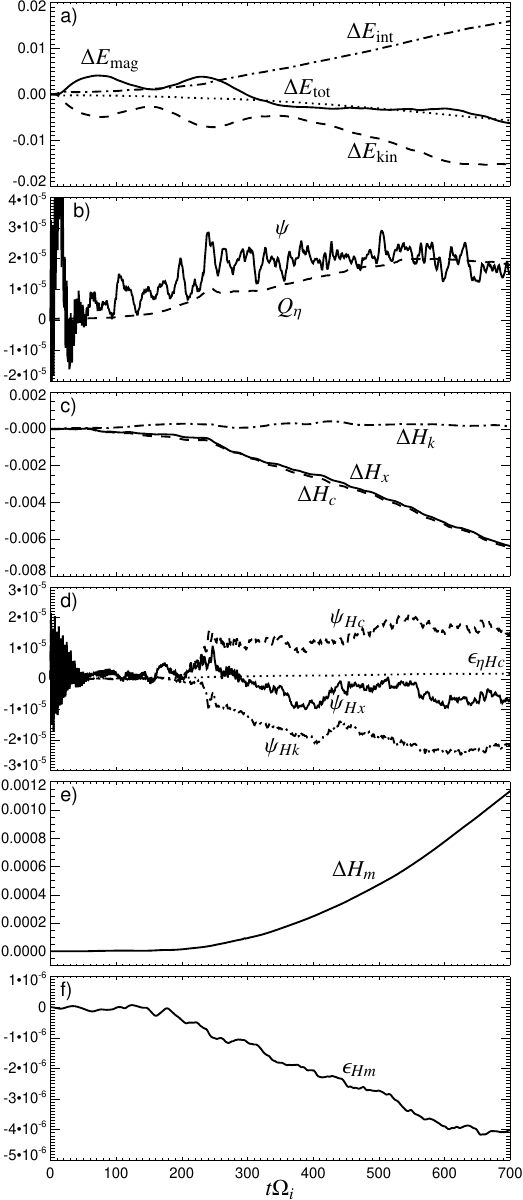}}
\caption{Evolution of different quantities as a function of time: (a) Relative changes
in the kinetic energy $\Delta E_{\text{kin}}$ (dashed line), magnetic energy  $\Delta E_{\text{mag}}$
(solid line), internal energy $\Delta E_{\text{int}}$ (dash-dotted line),
and total energy $\Delta E_{\text{tot}}$ (dotted line),
(b) resistive dissipation rate $Q_\eta$ (dashed line) and 
pressure-strain effective dissipation rate $\psi$ (solid line),
(c) relative changes in the kinetic helicity $\Delta H_k$ (dash-dotted line),
mixed helicity $\Delta H_x$ (solid line), and cross helicity $\Delta H_c$ (dashed line),
(d) resistive cross-helicity dissipation rate $\epsilon_{\eta Hc}$ (dotted line),
pressure-strain  effective cross-helicity dissipation rate $\psi_{Hc}$ (dashed line),
pressure-strain effective mixed-helicity dissipation rate $\psi_{Hx}$ (solid line), and pressure-strain effective kinetic-helicity dissipation rate $\psi_{Hk}$ (dash-dotted line), (e) relative change in the magnetic helicity  $\Delta H_m$, and
(f) resistive magnetic-helicity dissipation rate $\epsilon_{Hm}$.
\label{evol}
}
\end{figure}

Figure~\ref{evol} shows the evolution of different quantities as a function of time. 
In the simulation, the proton kinetic energy $E_{\text{kin}}$ and
the magnetic energy $E_{\text{mag}}$ oscillate with opposite
phases, suggesting energy exchanges, but overall these two quantities decrease.
On the other hand, the proton internal energy, $E_{\text{int}}=3\langle p \rangle/2,$
increases (here $p$ is the scalar pressure, $p=\mathrm{tr} (\mathbf{P})$,  $\mathrm{tr}$ being the trace). This heating (energisation) is weak initially ($t\Omega_i \lesssim 200$) and gets progressively
stronger as turbulence develops.
The total energy $E_{\text{tot}}=E_{\text{kin}}+E_{\text{int}}+E_{\text{mag}}$
slowly decreases owing to resistive dissipation because electrons are assumed to be 
massless and isothermal.
This behaviour is seen in Fig.~\ref{evol}a, which displays the relative changes
in the kinetic ($\Delta E_{\text{kin}}$), magnetic ($\Delta E_{\text{mag}}$),
internal ($\Delta E_{\text{int}}$), and total ($\Delta E_{\text{tot}}$) energies.
The proton energisation is driven by the pressure-strain coupling,  which plays the role
of an effective dissipation channel. Figure~\ref{evol}b displays 
the pressure-strain effective dissipation rate $\psi$ and
 the resistive dissipation rate $Q_\eta$ (dashed line).
The pressure-strain rate $\psi$ exhibits a large initial surge owing to
a relaxation of the initial conditions and then slowly increases and saturates;
$\psi$ oscillates in time largely owing to  
compressive effects. The resistive dissipation rate $Q_\eta$ slowly increases,
exhibits a local maximum at around $t\Omega_i \simeq 240$ (due to the
onset of reconnection).  The resistive dissipation rate reaches the maximum
at $t\Omega_i \simeq 550$ and slowly decreases thereafter.

The cross helicity $H_c$ decreases with time,
initially slowly, and later ($t\Omega_i \gtrsim 200$) the decrease
is faster and approximately linear in time; the kinetic helicity $H_k$ very slightly increases at the beginning
(from its nearly zero initial value)
and remains roughly constant later on ($t\Omega_i \gtrsim 200$); consequently,
the mixed helicity $H_x=H_c+H_k$  closely follows the evolution of the cross helicity $H_c$.
This is shown in Fig.~\ref{evol}c, which displays the evolution of 
the relative changes in the cross ($\Delta H_c$), mixed ($\Delta H_x$),
and kinetic ($\Delta H_k$) helicities.
We expected this evolution of the different helicities to be driven by the respective (effective) 
dissipation rates.  Figure~\ref{evol}d shows
the resistive cross-helicity dissipation rate $\epsilon_{\eta Hc}$, 
and the equivalents of the pressure-strain effective dissipation:
the cross-helicity rate $\psi_{Hc}$,
the kinetic-helicity rate $\psi_{Hk}$, and
the  mixed-helicity rate $\psi_{Hx}$.
The resistive cross-helicity dissipation rate is negligible 
but the cross-helicity pressure-strain rate $\psi_{Hc}$ is strong and
reaches a roughly constant value at  $t\Omega_i \gtrsim 240$, which is compatible
with the rate of  decrease in the cross helicity. We interpret 
$\psi_{Hc}$ as an effective dissipation rate of the cross helicity.
On the other hand, the kinetic-helicity $\psi_{Hk}$ and
the  mixed-helicity $\psi_{Hx}$ rates become negative for later times
($t\Omega_i \gtrsim 300$), which is at variance with the
time evolution of $H_k$ and $H_x$; interpretations of the terms
$\psi_{Hk}$ and $\psi_{Hx}$ are unclear.
The pressure-strain terms $\psi_{Hc}$ and $\psi_{Hk}$ are
dominated by incompressive fluctuations, and their compressive components
$\langle p(\boldsymbol{B}\cdot \boldsymbol{\nabla})\rho^{-1}\rangle$
and $\langle p (\boldsymbol{\omega}\cdot \boldsymbol{\nabla})\rho^{-1}\rangle$
are negligible due to small density variations 
(here $p$ is the scalar pressure, $p=\mathrm{tr} (\mathbf{P})$,  $\mathrm{tr}$ being the trace).
The combined energy and the cross helicity decrease
in time but the latter process is slower; consequently
the relative cross helicity increases with time.

Finally, the (modified) magnetic helicity increases with time 
in a quadratic manner as displayed in Fig.~\ref{evol}e,
which shows the relative change in this quantity ($\Delta H_m$). 
This evolution is driven by
the resistive magnetic-helicity dissipation rate $\epsilon_{Hm}$,
which becomes negative and decreases approximately linearly for $t\Omega_i \gtrsim 200$
(see Fig.~\ref{evol}f). 
In summary, Figure~\ref{evol} indicates that, for $t\Omega_i \gtrsim 550,$ the 
different dissipation rates are relatively constant, a property which is expected in a fully developed turbulence
system. Once this state is attained in a free decaying simulation, temporal
variations of the fluctuating quantities can be considered negligible (we test this
expectation using the KHM equations in section~\ref{KHMs}).

\begin{figure}
\centerline{\includegraphics[width=8.4cm]{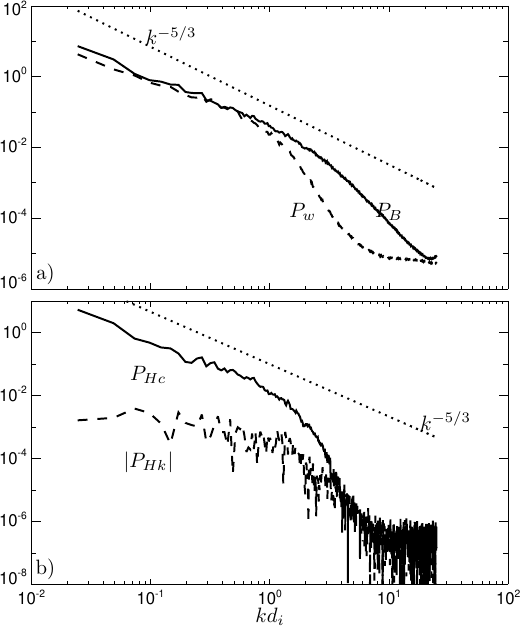}}
\caption{Omnidirectional spectral properties of different quantities at $t\Omega_i = 700$:
(a) Power spectral densities of (solid) the magnetic field $\boldsymbol{B}$, 
$P_B$ (solid), and 
compensated proton velocity field $\boldsymbol{w}$, $P_w$ (dashed), as a function of $k$
normalised to $d_i$.
(b) Cross helicity co-spectrum $P_{Hc}$ (solid) and
(absolute value of) kinetic helicity co-spectrum 
 $P_{Hk}$ (dashed) as a function of $k$.
The dotted lines show a spectrum $\propto k^{-5/3}$ for comparison.
\label{isospec}
}
\end{figure}

During the evolution, the kinetic and magnetic energies spread over a wide range of scales
and similar evolution is observed for the cross helicity.
Figure~\ref{isospec} shows their omnidirectional spectral properties at $t\Omega_i = 700$:
Figure~\ref{isospec}a displays power spectral densities of $\boldsymbol{B}$  and $\boldsymbol{w}$
fields, $P_B$ and $P_w$, respectively. These power spectral densities
exhibit a power-law-like behaviour at large scales 
(with a slope somewhat less steep than $-5/3$) and they
steepen at small scales, $P_w$ at around $k d_i \simeq 1$ and $P_B$ at around $k d_i \simeq 3$.
The kinetic spectral density $P_w$ becomes dominated by the noise for about $k d_i \gtrsim 6,$
whereas the magnetic spectral density  $P_B$ seems to be affected by the noise for $k d_i \gtrsim 20$. 

Figure~\ref{isospec}b shows co-spectra of the cross helicity $P_{Hc}$
and the (absolute value of the) kinetic helicity $P_{Hk}$.
$P_{Hc}$ exhibits a power-law like behaviour at large scales
with a slope similar to that of the energy power spectral densities and steepens
somewhere between $k d_i \simeq 1$ and $k d_i \simeq 3$, which are
the spectral break points of $P_w$ and $P_B$. 
The kinetic-helicity co-spectrum  $P_{Hk}$ oscillates between
positive and negative values and constitutes a small fraction of
$P_{Hc}$ at large scales and at small scales $P_{Hk}$ becomes comparable
to $P_{Hc}$; for $k d_i \gtrsim 6,$ both quantities are dominated by
the noise. As the magnetic helicity remains small during the whole simulation,
we did not investigate its spectral or spatial decomposition properties. 

\section{The K\'arm\'an-Howarth-Monin equations}
\label{KHMs}

To understand cross-scale transfers (cascades), exchanges,
and dissipations of energies and helicities,
we analysed the hybrid simulation results using the corresponding KHM equations.
We started with the combined energy.

\subsection{Energy}
\label{KHM1}
We characterised
the spatial scale decomposition of the kinetic and magnetic (and their sum) energies using
 the second-order structure functions:
\begin{equation*}
{S\!}_{w} =\frac{1}{4}\left\langle |\delta\boldsymbol{w}|^2\right\rangle, \ \
{S\!}_{B}= \frac{1}{4} \left\langle |\delta\boldsymbol{B}|^{2}\right\rangle
, \ \ \text{and}   \ \
S={S\!}_{w}+{S\!}_{B},
\end{equation*}
where the deltas denote increments of the corresponding quantities; for example,
$\delta\boldsymbol{w}=\boldsymbol{w}(\boldsymbol{x}+\boldsymbol{l})-\boldsymbol{w}(\boldsymbol{x})$,
with $\boldsymbol{x}$ being the position and
$\boldsymbol{l}$  the separation (lag) vector.
For the second-order structure function, $S,$
we obtain \citep{hellal24}
\begin{equation}
\partial_t S = K_\text{MHD} +K_\text{Hall} - \varPsi - D,
\label{KHM}
\end{equation}
where
$K_\text{MHD}$ and $K_\text{Hall}$ are the MHD and Hall 
non-linear cross-scale transfer (cascade) rates, respectively,
$\varPsi$ represents the pressure-strain effect, and
$D$ accounts for the effects of resistive dissipation.
We expressed 
the cascade rates $K_\text{MHD}$ and $K_\text{Hall}$ (as well as $\varPsi$ and $D$)
in the following form:

\begin{align}
K_\text{MHD}&=-\frac{1}{2}
\left\langle \delta \boldsymbol{w}\cdot \delta \left[ \left(\boldsymbol{u}\cdot\boldsymbol{\nabla}\right)\boldsymbol{w}
+\frac{1}{2}\boldsymbol{w}(\boldsymbol{\nabla}\cdot\boldsymbol{u}) \right] \right\rangle \nonumber \\
&-\frac{1}{2}
\left\langle \delta \boldsymbol{w}\cdot \delta \left(  \frac{\boldsymbol{J}\times\boldsymbol{B}}{\sqrt{\rho}}
\right) \right\rangle + \frac{1}{2} \left\langle  \delta \boldsymbol{J}\cdot
\delta \left(\boldsymbol{u}\times\boldsymbol{B}\right)
\right\rangle \\
K_\text{Hall}&= -\frac{1}{2}
\left\langle  \delta \boldsymbol{J}\cdot
\delta
\left(\boldsymbol{j}\times\boldsymbol{B}\right)
\right\rangle \\
\varPsi &= - \frac{1}{2} \left \langle \delta \boldsymbol{w} \cdot
\delta \left(\frac{\boldsymbol{\nabla}\cdot \mathbf{P}}{\sqrt{\rho}}\right)\right\rangle
\label{varpsi}
\\
D  &= \frac{1}{2} \eta \left \langle |\delta \boldsymbol{J}|^2\right \rangle = Q_\eta - \frac{1}{2} \eta \nabla^2 {S\!}_B.
\end{align}

\begin{figure}
\centerline{\includegraphics[width=8.4cm]{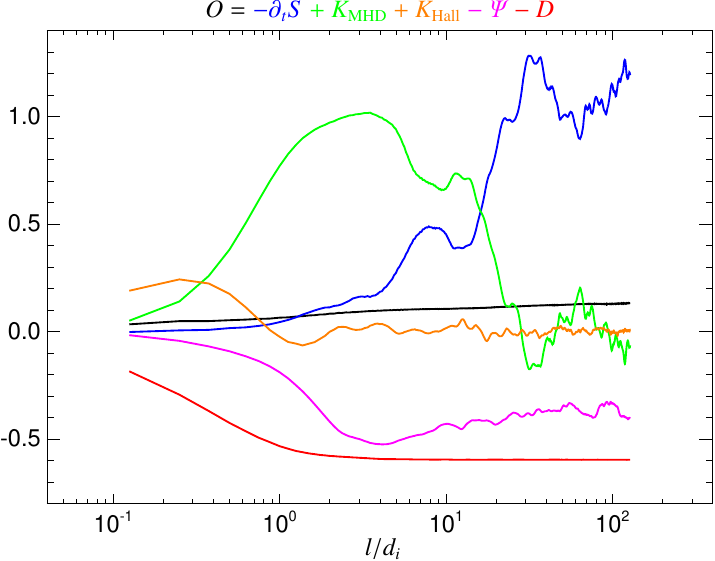}}
\caption{Isotropised energy KHM analysis at $t\Omega_i=700$: Validity test
$O$ (black line) as a function of $l$ along with the  different contributing terms:
decay rate $-\partial_t S$ (blue), MHD cascade rate $K_\text{MHD}$ (green),
Hall cascade rate  $K_\text{Hall}$ (orange), resistive dissipation rate $-D$ (red), and
pressure-strain rate $-\varPsi$ (magenta). All the quantities are normalised 
to the effective total dissipation rate $Q$.
\label{yag}
}
\end{figure}

The KHM equation constitutes a cross-scale energy conservation equation; we define the validity test $O$ as
\begin{equation}
O=-\partial_t S + K_\text{MHD} +K_\text{Hall} - \varPsi - D,
\label{KHMerr}
\end{equation}
which measures the error of the code owing to numerical issues.
Figure~\ref{yag} displays the isotropised energy KHM analysis
of the simulation results at the end of the simulation, $t\Omega_i=700$.
Figure~\ref{yag} shows the validity test $O$ and
the different contributing terms, $-\partial_t S\!$, $K_\text{MHD}$, $K_\text{Hall}$,  $- \varPsi$, and $-D$ 
(normalised to the total effective dissipation rate $Q$)
as a function of the scale separation $l=|\boldsymbol{l}|$ (normalised to $d_i$).
The KHM equation (\ref{KHM}) is relatively well satisfied ($|O|/Q\lesssim 10$ \%);
this error is connected with the noise level owing to the particle-in-cell scheme.
The energy decay $\partial_t S\!$ is strong at large scales 
$l\gtrsim 10 d_i$, but becomes less important and is negligible somewhat below $l=d_i$. 
The MHD cascade rate is around zero at large scales and becomes important
at intermediate scales ($ 6 d_i \gtrsim l\gtrsim 1 d_i$),
where it reaches a maximum value of about the dissipation rate $Q$.
At small scales, part of the cascade
continues via the Hall term and  $K_\text{Hall}$ attains a
maximum value of about $20 \%$ of $Q$; at these scales, 
the resistive dissipation and the pressure-strain interaction are also active.

\begin{figure}
\centerline{\includegraphics[width=8.4cm]{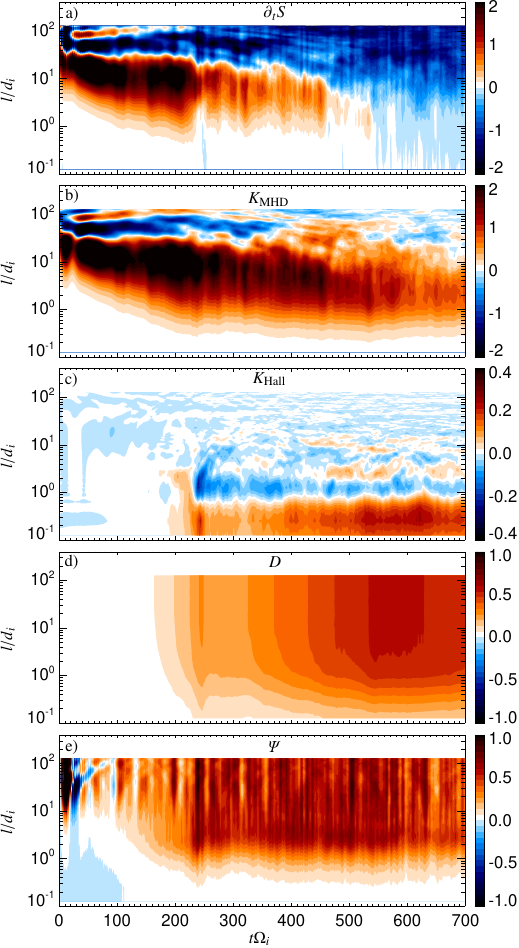}}
\caption{Evolution of isotropised energy KHM results.
Shown are the different
KHM terms as a function of time $t$ and $l$:
(a) decay rate $\partial_t S$,
(b) MHD cascade rate $K_\text{MHD}$,
(c) Hall cascade rate  $K_\text{Hall}$,
(d) resistive dissipation rate $D$, and
(e) pressure-strain rate $\varPsi$.
All the quantities are normalised to the effective total dissipation
rate $Q$ (averaged over $600 \le t\Omega_i \le 700$).
\label{evolKHM}
}
\end{figure}

Figure~\ref{yag} shows the properties of well-developed decaying turbulence
\cite[see Fig. 3 of][]{hellal24}.
The combined energy decays at large scales, cascades at
intermediate scales, and dissipates at small scales.
As the KHM equation remains relatively well satisfied
throughout the entire simulation, including the turbulence onset, we can analyse
the temporal evolution of each KHM term at all times. We
can determine which term or process dominates (at a given scale) during the onset 
and once the cascade is fully developed.
Figure~\ref{evolKHM} shows the evolution of the isotropised KHM results,
the different contributing terms
as a function of time, and the separation scale $l$
 normalised to the total effective dissipation rate $Q$ (averaged over
the time interval $600 \le t \Omega_i \le 700$).
Figure~\ref{evolKHM} demonstrates that the features of well-developed
turbulence seen in Fig.~\ref{yag} only appear at later times.
During an initial phase ($0 \le t \Omega_i \le 200$), the system  
is governed by $\partial_t S \simeq K_\text{MHD}$, and
increasingly small scales are generated through 
the (MHD) non-linear coupling as $\partial_t S$ is positive at intermediate scales.
At later times (for about $ t \Omega_i \lesssim 550$), 
$\partial_t S$ becomes negative,  monotonically
decreasing with $l$ as expected.
As the energy transfer towards smaller scales continues,
the resistive and the effective pressure-strain dissipation 
effects gradually appear 
 (their behaviour corresponds to the evolution
of the resistive $Q_\eta$ and pressure-strain rates $\psi$ in Fig.~\ref{evol}b). 
The Hall term gradually sets in
around $180 \lesssim t \Omega_i \lesssim 230$.

\subsection{Cross and kinetic helicities}
\label{KHM2}

Next we continued with the cross, kinetic, and mixed helicities.
First, we characterised the cross helicity using a second-order structure function \cite[cf.][]{baga16,baga17} as
\begin{equation}
{S\!}_{Hc} =\frac{1}{2}\left\langle \delta\boldsymbol{u} \cdot \delta \boldsymbol{B}  \right\rangle.
\end{equation}
For ${S\!}_{Hc}$ we get the following dynamic KHM equation:
\begin{equation}
\partial_{t}{S\!}_{Hc}=K_{\text{MHD}Hc}+K_{\text{Hall}Hc}-\varPsi_{Hc}-D_{Hc},
\label{KHM_Hc}
\end{equation}
where 
\begin{align}
K_{\text{MHD}Hc}&=-\frac{1}{2}\left\langle \delta\boldsymbol{B}\cdot\delta\left[(\boldsymbol{u}\cdot\boldsymbol{\nabla})\boldsymbol{u}\right]\right\rangle 
+\frac{1}{2}\left\langle \delta\boldsymbol{B}\cdot\delta\left(\frac{\boldsymbol{J}\times\boldsymbol{B}}{\rho}\right)\right\rangle \nonumber \\
&+\frac{1}{2}\left\langle \delta\boldsymbol{\omega}\cdot\delta\left(\boldsymbol{u}\times\boldsymbol{B}\right)\right\rangle  \\
K_{\text{Hall}Hc}&=-\frac{1}{2}\left\langle \delta\boldsymbol{\omega}\cdot\delta\left(\boldsymbol{j}\times\boldsymbol{B}\right)\right\rangle \\
D_{Hc}&=\frac{\eta}{2}\left\langle \delta\boldsymbol{\omega}\cdot\delta\boldsymbol{J}\right\rangle \\
\varPsi_{Hc}&=\frac{1}{2}\left\langle \delta\boldsymbol{B}\cdot\delta\left(\frac{\boldsymbol{\nabla}\cdot\boldsymbol{\mathrm{P}}}{\rho}\right)\right\rangle.
\end{align}
Here, 
$K_{\text{MHD}Hc}$ and $K_{\text{Hall}Hc}$ represent the MHD and Hall
non-linear coupling terms, respectively,
$D_{Hc}$ accounts for the effects of resistive dissipation, whereas
$\varPsi$ characterises the cross-helicity equivalent of the pressure-strain effect.

Similarly, we define a second-order structure function for the kinetic helicity:
\begin{equation}
{S\!}_{Hk} =\frac{m}{4e}\left\langle \delta\boldsymbol{u} \cdot \delta \boldsymbol{\omega} \right\rangle.  
\end{equation}
For ${S\!}_{Hk}$, the KHM equation reads 
\begin{equation}
\partial_{t}{S\!}_{Hk}=K_{\text{MHD}Hk}+K_{\text{Hall}Hk}-\varPsi_{Hk}
\label{KHM_Hk}
,\end{equation}
where
\begin{align}
K_{\text{MHD}Hk}&=-\frac{m}{2e}\left\langle \delta\boldsymbol{\omega}\cdot\delta\left[(\boldsymbol{u}\cdot\boldsymbol{\nabla})\boldsymbol{u}\right]\right\rangle  ,\\
K_{\text{Hall}Hk}&=-K_{\text{Hall}Hc} ,\\
\varPsi_{Hk}&=\frac{m}{2e}\left\langle \delta\boldsymbol{\omega}\cdot\delta\left(\frac{\boldsymbol{\nabla}\cdot\boldsymbol{\mathrm{P}}}{\rho}\right)\right\rangle.
\end{align}
Here, analogously to the cross helicity,
$K_{\text{MHD}Hk}$ and $K_{\text{Hall}Hk}$ represent the MHD and Hall
non-linear coupling terms, respectively, and
$\varPsi_{Hk}$ characterises the kinetic-helicity equivalent of the pressure-strain  effect.

Finally, for the mixed helicity, we define a second-order structure function as
a sum of the two previous ones,
\begin{equation}
{S\!}_{Hx}={S\!}_{Hc} + {S\!}_{Hk},
\end{equation}
and we obtain the corresponding KHM equation (again as a sum of the two previous ones):  
\begin{equation}
\partial_{t}{S\!}_{Hx}=K_{\text{MHD}Hx}-\varPsi_{Hx}-D_{Hc},
\label{KHM_Hx}
\end{equation}
where 
\begin{align}
K_{Hx}&= K_{\text{MHD}Hc}+ K_{\text{MHD}Hk} \\
\varPsi_{Hx}&=\varPsi_{Hc}+ \varPsi_{Hk}.
\end{align}
In the mixed-helicity KHM equation, $K_{Hx}$ represents the MHD 
non-linear coupling term (the Hall contribution cancels out),
$D_{Hc}$ remains the cross-helicity resistive term,
 and $\varPsi_{Hx}$ characterises the mixed-helicity equivalent of the pressure-strain
effect.

\begin{figure}
\centerline{\includegraphics[width=8.4cm]{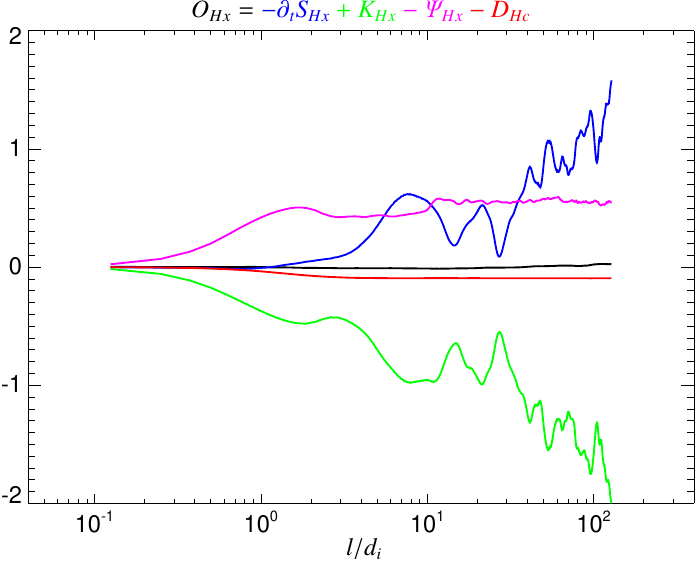}}
\caption{
Isotropised mixed-helicity KHM analysis at $t\Omega_i=700$:
 Validity test
$O_{Hx}$ (black line) as a function of $l$ 
along with the  different contributing terms:
decay rate $-\partial_t {S\!}_{Hx}$ (blue),
MHD term $K_{Hx}$ (green),
resistive dissipation rate $-D_{Hc}$ (red), and
pressure-strain rate $-\varPsi_{Hx}$ (magenta).
All the quantities are normalised to the effective total (cross-helicity) dissipation
rate $\epsilon_{Hc}$.
\label{hxyag}
}
\end{figure}

Starting with the mixed helicity, we define the validity test $O_{Hx}$,
\begin{equation}
O_{Hx}=-\partial_{t}{S\!}_{Hx}+K_{Hx}-\varPsi_{Hx}-D_{Hc}
\label{KHMerrHx}
,\end{equation}
as in the energy case.
Figure~\ref{hxyag} displays the isotropised mixed-helicity KHM analysis
of the simulation results at the end of the simulation, $t\Omega_i=700$,
showing the validity test $O_{Hx}$ and
the different contributing terms ($-\partial_t S\!_{Hx}$, $K_{Hx}$,  $- \varPsi_{Hx}$, $-D_{Hc}$)
as a function of the scale separation $l=$.
These quantities are normalised to the total effective dissipation rate  $\epsilon_{Hc}=\epsilon_{\eta Hc}+\psi_{Hc}$, 
a sum of of the cross-helicity resistive and pressure-strain rates. 
As for the energy, the mixed-helicity KHM Eq.~(\ref{KHM_Hx}) 
is relatively well satisfied ($|O_{Hx}|/\epsilon_{Hc}\lesssim 5$ \%).

Figure~\ref{hxyag} exhibits a very different behaviour from that seen in Fig.~\ref{yag} for the energy. The decay term
$\partial_t S\!_{Hx}$ is important (and negative) at large
scales. The dissipation term $D_{Hc}$ is positive but small, and
the resistive dissipation of cross and mixed helicities is weak.
While $\partial_t S\!_{Hx}$ and $D_{Hc}$ behave similarly to the their
energy counterparts, $K_{Hx}$ and $\varPsi_{Hx}$ are of opposite sign.
The mixed-helicity pressure-strain term $\varPsi_{Hx}$ has a negative sign, 
which suggests that the pressure-strain coupling generates the mixed helicity at small
scales. The non-linear term $K_{Hx}$  also has a negative sign, which may indicate
an inverse cascade of the mixed helicity. These two suggestions are at variance with
the overall decrease in the mixed helicity; consequently, we cannot 
interpret $K_{Hx}$ as the cascade rate and
 $\varPsi_{Hx}$ as the effective dissipation rate of the mixed helicity.

\begin{figure}
\centerline{\includegraphics[width=8.4cm]{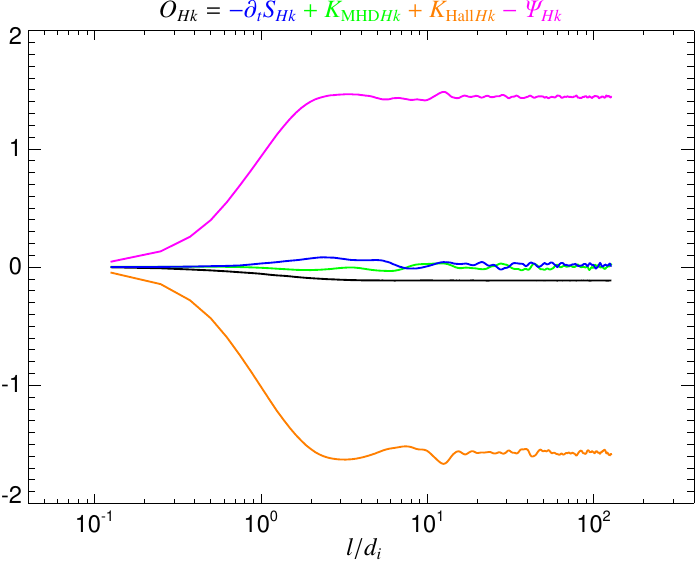}}
\caption{
Isotropised kinetic-helicity KHM analysis at $t\Omega_i=700$:
 Validity test
$O_{Hk}$ (black line) as a function of $l$
along with the  different contributing terms:
the decay rate $-\partial_t {S\!}_{Hk}$ (blue),
the MHD cascade rate $K_{\text{MHD}Hk}$ (green),
the Hall cascade rate $K_{\text{Hall}Hk}$ (orange), and
the pressure-strain rate $-\varPsi_{Hk}$ (magenta).
All the quantities are normalised to the effective total (cross-helicity) dissipation
rate $\epsilon_{Hc}$.
\label{hkyag}
}
\end{figure}

To better understand the behaviour of the mixed helicity, we looked at
its two contributions, $H_c$ and $H_k$, separately. 
We started with the kinetic helicity KHM equation~(\ref{KHM_Hk}),
defining the corresponding validity test $O_{Hk}$,
\begin{equation}
O_{Hk}=-\partial_{t}{S\!}_{Hk}+K_{\text{MHD}Hk}+K_{\text{Hall}Hk}-\varPsi_{Hk}.
\label{KHMerrHk}
\end{equation}
Figure~\ref{hkyag} displays the isotropised kinetic-helicity KHM analysis
of the simulation results at the end of the simulation, $t\Omega_i=700$;
it shows the validity test $O_{Hk}$ and
the different contributing terms, $-\partial_t S\!_{Hk}$, $K_{\text{MHD}Hx}$, $K_{\text{Hall}Hx}$, and 
 $- \varPsi_{Hk}$ (normalised to the total effective dissipation rate  $\epsilon_{Hc}$)
as a function of the scale separation $l$.
The kinetic-helicity KHM equation is well satisfied ($|O_{Hk}|/\epsilon_{Hc}\lesssim 10$ \%),
but the behaviours of the different terms are very different
from the energy KHM results in Fig.~\ref{yag}.
The decay term $\partial_t S\!_{Hk}$ and the MHD coupling term $K_{\text{MHD}Hx}$
are negligible; this is in agreement with the evolution of the kinetic helicity,
which is roughly constant. In contrast, the Hall coupling term $K_{\text{Hall}Hx}$
and the pressure-strain term $\varPsi_{Hk}$ are strong and mostly compensate
each other, 
\begin{equation}
K_{\text{Hall}Hk}\simeq\varPsi_{Hk}.
\end{equation}
Therefore, the Hall term couples to the kinetic-helicity 
pressure-strain term.

\begin{figure}
\centerline{\includegraphics[width=8.4cm]{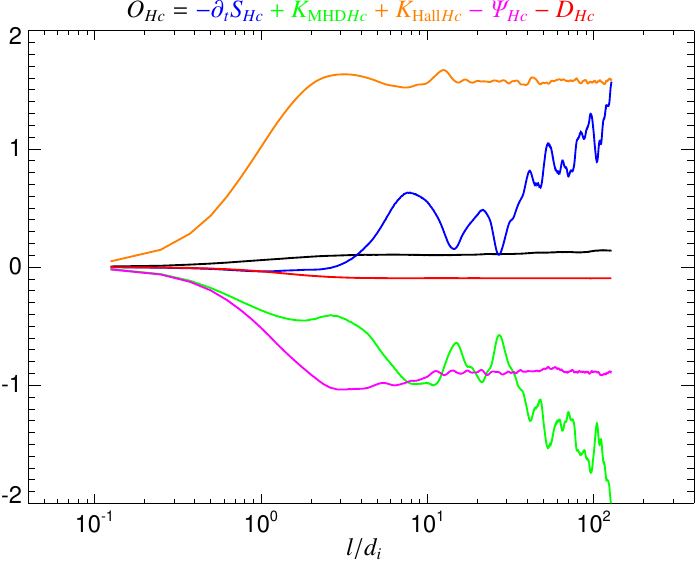}}
\caption{
Isotropised cross-helicity KHM analysis at $t\Omega_i=700$:
 Validity test $O_{Hc}$ (black line) as a function of $l$
along with the  different contributing terms:
the decay rate $-\partial_t {S\!}_{Hc}$ (blue),
the MHD term $K_{\text{MHD}Hc}$ (green),
the Hall term $K_{\text{Hall}Hc}$ (orange),
the resistive dissipation rate $-D_{Hc}$ (red), and
the pressure-strain rate $-\varPsi_{Hc}$ (magenta).
All the quantities are normalised to the effective total (cross-helicity) dissipation
rate $\epsilon_{Hc}$.
\label{hcyag}
}
\end{figure}

We continued with the cross-helicity KHM equation~(\ref{KHM_Hc}):
we define the corresponding validity test $O_{Hc}$,
\begin{equation}
O_{Hc}=-\partial_{t}{S\!}_{Hc}+K_{\text{MHD}Hc}+K_{\text{Hall}Hc}-\varPsi_{Hk} - D_{Hc}
\label{KHMerrHc}
.\end{equation}
Figure~\ref{hcyag} displays the isotropised cross-helicity KHM analysis
of the simulation results at the end of the simulation, $t\Omega_i=700$;
it shows the validity test $O_{Hc}$ and
the different contributing terms, $-\partial_t S\!_{Hc}$, $K_{\text{MHD}Hc}$, $K_{\text{Hall}Hc}$,
 $- \varPsi_{Hc}$, and $- D_{Hc}$, as a function of the scale separation $l$.
The cross-helicity KHM equation is relatively well satisfied ($|O_{Hc}|/\epsilon_{Hc}\lesssim 15$ \%).
The results of Fig.~\ref{hcyag} become somewhat similar to those of 
the energy KHM results of Fig.~\ref{yag}: the decay term
$\partial_t S\!_{Hc}$ is negative and dominates at large scales; and
the resistive  and the cross-helicity pressure-strain rates
 behave as dissipation channels and are important at small scales. However,
the MHD and Hall coupling terms remain strange. 
Nevertheless, when we combine them into one term, $K_{Hc}=K_{\text{MHD}Hc}+K_{\text{Hall}Hc}$,
the resulting quantity exhibits an  analogous behaviour to the energy cascade rate (see Fig.~\ref{yag}).
\begin{figure}
\centerline{\includegraphics[width=8.4cm]{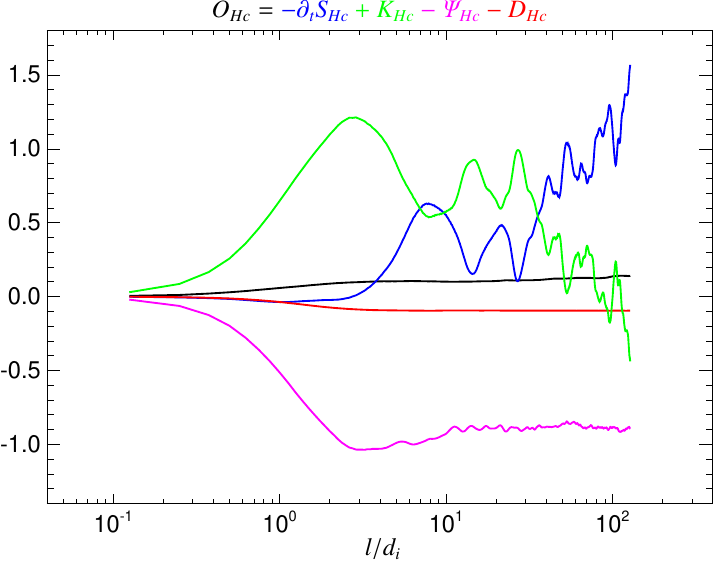}}
\caption{
Rearranged isotropised cross-helicity KHM analysis of Fig.~\ref{hcyag}: Validity test
$O_{Hc}$ (black line),
decay rate $-\partial_t {S\!}_{Hc}$ (blue),
MHD+Hall term $K_{Hc}$ (green),
resistive dissipation rate $-D_{Hc}$ (red), and
pressure-strain rate $-\varPsi_{Hc}$ (magenta) 
as a function of $l$.
All the quantities are normalised to the effective total (cross-helicity) dissipation
rate $\epsilon_{Hc}$.
\label{hcyag2}
}
\end{figure}
Figure~\ref{hcyag2} repeats the results of Fig.~\ref{hcyag} but combines
the MHD and Hall contributions into one term, $K_{Hc}$. This indicates
that $K_{Hc}$ describes the total (MHD+Hall) cascade rate of the cross helicity.
\begin{figure}
\centerline{\includegraphics[width=8.4cm]{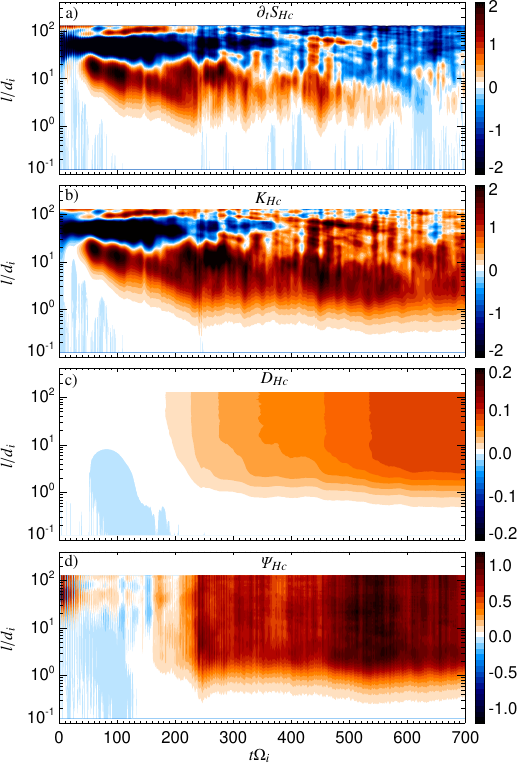}}
\caption{Evolution of isotropised cross-helicity KHM results.
Shown are the different
KHM terms as a function of time $t$ and $l$:
(a) decay rate $\partial_t S\!_{Hc}$,
(b) MHD+Hall cascade rate $K_{Hc}$,
(c) resistive dissipation rate $D_{Hc}$, and
(d) pressure-strain rate $\varPsi_{Hc}$.
All the quantities are normalised to the effective total dissipation
rate $\epsilon_{Hc}$ (averaged over $600 \le \Omega_i \le 700$).
\label{evolKHM_Hc}
}
\end{figure}
Figure~\ref{hcyag2} indicates that the cross helicity decays at
large scales, cascades  at intermediate scales, and dissipates
at small scales via the resistive and effective pressure-strain 
dissipation effects, at least at the end of simulation. We used
the cross-helicity KHM equation to look at the development of the system.
Figure~\ref{evolKHM_Hc} shows the evolution of the isotropised KHM results,
the different contributing terms
as a function of time, and the separation scale $l$
 normalised to the total effective dissipation rate $\epsilon_{Hc}$ (averaged over
the time interval $600 \le t \Omega_i \le 700$).
Figure~\ref{evolKHM_Hc} reveals a very similar behaviour to
the energy properties in Fig.~\ref{evolKHM}.
During the initial phase ($0 \le t \Omega_i \le 200$), the system
is governed by $\partial_t {S\!}_{Hc} \simeq K_{Hc}$,
and increasingly small scales are generated through
the (MHD) non-linear coupling as $\partial_t {S\!}_{Hc}$ is positive at intermediate scales.
At later times (for about $ t \Omega_i \lesssim 550$), 
$\partial_t {S\!}_{Hc}$ becomes mostly negative and negligible 
at small scales. 
As the energy transfer towards smaller scales continues,
the resistive dissipation and more importantly the
pressure-strain effective dissipation gradually appear
 (corresponding to the evolution 
of the resistive $\epsilon_{\eta Hc}$ and pressure-strain rates $\psi_{Hc}$ 
in Fig.~\ref{evol}d).

\section{Discussion}
\label{discuss}
In this paper we present an investigation of 2D decaying plasma turbulence 
in a weakly collisional plasma with an out-of-plane background magnetic field
using a pseudo-spectral hybrid code.
The plasma system is initialised with substantial fluctuating kinetic and
magnetic energies as well as cross helicity at relatively large scales; these initial
fluctuations exhibit no overall magnetic or kinetic helicities.
We analyse the simulation results using the KHM equation 
for the combined (kinetic plus magnetic) energy, and cross, kinetic, and mixed helicities.
These different KHM equations, which characterise the cross-scale conservation
of the given quantity, are well satisfied. These code properties are partly attributable to 
the pseudo-spectral scheme, which improves the cross-scale  conservation properties
compared to a finite-difference scheme. 
The KHM results show that the combined energy, which is initially distributed at
large scales, transfers to smaller scales owing to the non-linear MHD term 
(and later the Hall one). After a transient period, a 
fully developed turbulent system arises,
where the energy decays at large scales, cascades at intermediate scales, and
is dissipated at small scales via the resistive term and the pressure-strain
term; this system plays the role of an effective dissipation mechanism. These
results agree with previous hybrid simulation results \citep{hellal22,hellal24}.

The properties of the cross, kinetic, and mixed helicities are more complicated. In the hybrid
approximation, we expected the mixed helicity (combination of the cross and kinetic helicities)
to be the relevant quantity, which decays, cascades, and dissipates. 
However, the simulation results show that the cross helicity
is the relevant quantity in weakly collisional plasmas. The corresponding KHM equation  
indicates that the cross helicity behaves analogously to the combined energy:
it decays at large scales, cascades via
the MHD and Hall non-linear coupling at intermediate scales, and dissipates 
at small scales via the resistive term and, more importantly,
via the cross helicity equivalent of the pressure-strain term, which again plays a role of an effective dissipation mechanism.

The combined energy and the cross helicity decrease
over time, but the latter process is slower, and consequently
the relative cross helicity increases.
This is in agreement with the theoretical incompressible MHD expectations
and numerical simulations \cite[e.g.][]{dobral80,montal22},
but is at variance with observations \citep{bavaal98}.
This effect may be related to large-scale gradients in the solar wind
due to its expansion
\citep{grapal22} and/or velocity shears \citep{robeal92}; 
we note that, 
assuming homogeneity, these effects can be included in the KHM equation \citep{wanal09,stawal11,hellal13}.

The magnetic helicity is a separate ideal invariant in the hybrid system
\cite[similarly to the Hall MHD case; see][]{poyo22}. In the simulation,
the  magnetic helicity is initially negligible and 
is scantily generated via the resistive term (as the magnetic helicity
is not a positively definite quantity, this term may lead to its production).
We did not observe any coupling between the magnetic
and cross helicities, in agreement with theoretical expectations.
We also derived and analysed the KHM equation for the magnetic
helicity; these results indicate that the magnetic helicity
does not exhibit any cascade (not shown here).
We investigated a system where the mixed and magnetic helicities are
not coupled, but in a more general situation it is necessary to
investigate the generalised (canonical) helicity,
that is, a combination of the mixed and magnetic helicities \cite[cf.][]{poyo22}.
In such a situation, there may be coupling between the cross and magnetic helicities,
which could lead to a reduction of the energy cascade, 
a phenomenon called helicity barrier \citep{meyral21}. 
\cite{squial23} interpreted some hybrid simulation results 
as being a consequence of the helicity barrier. Our results, however, show that 
the helicity barrier is not relevant in hybrid (or Hall MHD) systems.

In this work, we used the KHM equations to analyse numerical simulation results for 
the properties of the energy and the cross helicity. 
Based on the similarities between the KHM results for the two quantities,
we conclude that, even though the cross helicity is not
an ideal invariant, it is this quantity that decays,
cascades, and dissipates in parallel with the combined energy.
As our results could simply be due to some peculiar properties of the KHM
approach, we investigated analogous spectral-transfer and coarse-graining approaches
(see Appendices~\ref{appen} and~\ref{apphc}).
We obtained equivalent results for the energy and the cross helicity, allowing us to conclude that 
our results are robust.
However, we investigated only one 2D simulation with a 
particular set of parameters; more simulations are needed to study
the roles of the various plasma and turbulence parameters.
In particular, it would be interesting to check how different values of magnetic helicity  
(and its cascade) influence properties of plasma turbulence \cite[cf.][]{strial95}.
Furthermore, in the simulation, the initial
kinetic helicity was negligible and remained negligible during the whole simulation.
It is not clear what would happen if the simulation were to start with a strong kinetic helicity.
This work also needs to be extended to three dimensions to account 
for the anisotropy induced by the background magnetic field \citep{verdal15,montal22,hellal24}.
It is also necessary to study a fully kinetic regime;
the electron pressure strain effect may act at relatively large scales
for the energy \citep{yangal22,manzal24} and may influence
the evolution of the cross helicity. 

In conclusion, this paper we show that 
the dynamics of cross helicity in weakly collisional plasmas is strongly affected
over a wide range of scales (even well above ion scales) by the Hall and pressure-strain physics.
Observational tests of these results are challenging; the relevant quantities contain 
the vorticity field, which is impossible to measure with just one spacecraft. 
The incompressible MHD KHM equation is often used to estimate energy and cross-helicity cascade rates
from in situ spacecraft measurements in the solar wind  \citep{sorral07,macbal08,stawal09,maso23}.
Our results, along with those of previous studies, show that at large scales this approximation is sufficient for the energy;
the Hall correction appears at the ion scale \citep{bandal20b} and
the pressure-strain effect for the energy also becomes important at around this scale 
\citep{yangal22,manzal24}. On the other hand,
our results indicate that the MHD estimates of the cross-helicity cascade rate
are inadequate in the solar wind; 
in our simulation, the cross-helicity MHD cascade rate is weaker than the Hall cascade rate
and they even have the opposite sign.

\begin{acknowledgements}
This work was supported by the Ministry of Education, Youth and Sports of the Czech Republic through the e-INFRA CZ (ID:90254).
The authors thank T.~Tullio for a continuous enriching support and acknowledge useful
discussions with A. Verdini, S. Landi, L. Matteini, Emanuele Papini, and L. Franci.
\end{acknowledgements}

\begin{appendix}

\section{Energy}
\label{appen}

\subsection{Spectral transfer approach}
\label{appen1}

Another possibility how to quantify the turbulence properties is the spectral approach
\cite[cf.][]{minial07,gretal17}.
We characterised the combined energy by a low-pass filtered quantity \citep{hellal21b,hellal24}
\begin{align}
E_{k}&=\frac{1}{2}\sum_{|\boldsymbol{k}^\prime|\le k}\left(|\widehat{\boldsymbol{w}}|^{2}+|\widehat{\boldsymbol{B}}|^{2}\right),
\end{align}
where the wide hat denotes the Fourier transform.
For the spectrally-decomposed quantity $E_{k}$ we
 obtained the following dynamic equation
\begin{align}
\partial_t E_{k}+{S\!}_{\text{MHD}k}+{S\!}_{\text{Hall}k}=-\Psi_{k}-D_{k},
\label{sptrdyn}
\end{align}
where
${S\!}_{\text{MHD}k}$ and ${S\!}_{\text{Hall}k}$ represent the MHD and Hall energy transfer rates, respectively;
$\Psi_k$ describes the pressure-strain effect;
and $D_k$ is the resistive dissipation rate for modes with wave-vector magnitudes smaller than or equal to $k$.
They can be expressed as
\begin{align}
{S\!}_{\text{MHD}k} &=\Re\sum_{|\boldsymbol{k}^\prime|\le k} \bigg[ \widehat{\boldsymbol{w}}^{*}\cdot\widehat{(\boldsymbol{u}\cdot\boldsymbol{\nabla})\boldsymbol{w}}+\frac{1}{2}\widehat{\boldsymbol{w}}^{*}\cdot\widehat{\boldsymbol{w}(\boldsymbol{\nabla}\cdot\boldsymbol{u})} \nonumber \\
& \phantom{=}-\widehat{\boldsymbol{w}}^{*}\cdot\widehat{\rho^{-1/2}\boldsymbol{J}\times\boldsymbol{B}}
 -\widehat{\boldsymbol{B}}^{*}\cdot\widehat{\boldsymbol{\nabla}\times\left(\boldsymbol{u}\times\boldsymbol{B}\right)} \bigg],
\\
{S\!}_{\text{Hall}k} &=\Re \sum_{|\boldsymbol{k}^\prime|\le k} \widehat{\boldsymbol{B}}^{*}\cdot\widehat{\boldsymbol{\nabla}\times\left(\boldsymbol{j}\times\boldsymbol{B}\right)},
\\
\Psi_{k}&=\Re \sum_{|\boldsymbol{k}^\prime|\le k} \widehat{\boldsymbol{w}}^{*}\cdot\widehat{\rho^{-1/2}\boldsymbol{\nabla}\cdot \mathbf{P}},
\\
D_{k}&=\eta \sum_{|\boldsymbol{k}^\prime|\le k}  |\boldsymbol{k}^\prime|^{2}|\widehat{\boldsymbol{B}}|^{2}.
\end{align}
In these expression the asterisk denotes the complex
conjugate, and $\Re$ denotes the real part.
For fully developed turbulence
${S\!}_{\text{MHD}k}$ and ${S\!}_{\text{Hall}k}$ are the MHD and Hall cascade rates, respectively.

Similarly to the KHM case, we defined the validity test of the spectral transfer (ST) equation,
Eq.~(\ref{sptrdyn}), as
\begin{equation}
O_k=
\partial_t E_{k}+{S\!}_{\text{MHD}k}+{S\!}_{\text{Hall}k}+\Psi_{k}+D_{k}.
\label{ok}
\end{equation}
\begin{figure}[htb]
\centerline{
\includegraphics[width=7.8cm]{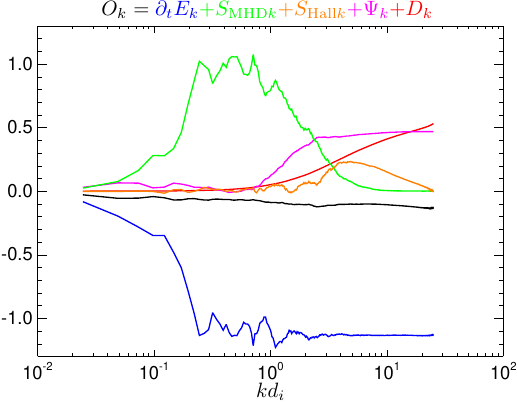}
}
\caption{
Spectral transfer (ST) analysis of the combined energy,
the validity test of the isotropic
ST $O_k$, Eq.~(\ref{ok}),  as a function of $k$
at $t \Omega_i = 700$ (black line)
along with
the different contributing terms:
  decay rate $\partial_t E_{k}$ (blue),
   MHD cascade rate ${S\!}_{\text{MHD}k}$ (green),
   Hall cascade rate ${S\!}_{\text{Hall}k}$ (orange),
 resistive dissipation rate $D_k$ (red), and
 pressure-strain rate $\Psi_k$ (magenta).
All the quantities are given in units of the total (effective) dissipation rate
$Q$.
\label{sptr}
}
\end{figure}
Figure~\ref{sptr} shows the ST validity test $O_k$ and
the contributing terms as functions of $k$ at $t \Omega_i = 700$
The ST equation (\ref{sptrdyn}) is relatively well satisfied ($|O_k|/Q\lesssim 14$ \%).
Figure~\ref{sptr} quantifies the turbulence properties, the energy decays at large
scales, cascades at intermediate scales, and at small scales the
cascade partly continues via the Hall term and partly is dissipated
by the resistive and pressure-strain effects.
The ST results in Fig.~\ref{sptr} are quantitatively equivalent to the corresponding KHM results (see Fig.~\ref{yag})
through the inverse
proportionality $ k l = \sqrt{2}$. Using this relationship 
we got
\begin{equation}
{S\!}_{\text{MHD}k} \simeq K_\text{MHD}, \  \
 {S\!}_{\text{Hall}k} \simeq K_\text{Hall}.
\end{equation}
The other terms exhibit a complementarity properties;
the ST approach  is by construction (spectral) low-pass filter
whereas the KHM approach exhibits rather high-pass (or spatial low-pass) filter
behaviour
\cite[see][]{hellal21b,hellal24}.

\subsection{Coarse-graining approach}

\label{appen2}
An alternative way how to analyse the turbulence properties
is the spatial filtering 
\cite[][]{eyal09,alui11,manzal22}.
In this approach the scale decomposition of the combined energy
is done using a spatial filter (coarse graining)
\begin{equation}
E_l=
\frac{1}{2}\left\langle \overline{\rho}|\widetilde{\boldsymbol{u}}|^{2}+|\overline{\boldsymbol{B}}|^{2}\right\rangle 
\end{equation}
where the line denotes a spatial filtering  
\begin{equation}
\overline{\boldsymbol{a}}_l(\boldsymbol{x})=\int\mathrm{d}^{2}x G_{l}(\boldsymbol{r})\boldsymbol{a}(\boldsymbol{x}+\boldsymbol{r}) 
\end{equation}
and the tilde denotes the corresponding density-weighted or  
Favre filtering \citep{favr69}
\begin{equation}
\widetilde{\boldsymbol{a}}_l=\frac{\overline{\rho\boldsymbol{u}}_l}{\overline{\rho}_l}.
\end{equation}
For the filtered energy we obtained this dynamic equation
\begin{equation}
\partial_t E_l + \Pi_{\text{MHD}l} +  \Pi_{\text{Hall}l} = -\Phi_l -\mathcal{D}_l
\label{eqcgen}
\end{equation}
where
$ \Pi_{\text{MHD}l}$ and $\Pi_{\text{Hall}l}$
represent the MHD and Hall energy transfer rates, respectively;
$\Phi_l$ and  $\mathcal{D}_l$ describe the pressure-strain and resistive dissipation
 effects, respectively. These terms can be expressed as
\begin{align}
 \Pi_{\text{MHD}l} &= \left\langle \boldsymbol{\nabla}\widetilde{\boldsymbol{u}}:\overline{\rho}\widetilde{\boldsymbol{\tau}}\right\rangle 
-\left\langle \widetilde{\boldsymbol{u}}\cdot\overline{\boldsymbol{J}\times\boldsymbol{B}}\right\rangle 
-\left\langle \overline{\boldsymbol{J}}\cdot\overline{\left(\boldsymbol{u}\times\boldsymbol{B}\right)}\right\rangle  \\
\Pi_{\text{Hall}l} &= \left\langle \overline{\boldsymbol{J}}\cdot\overline{\left(\boldsymbol{j}\times\boldsymbol{B}\right)}\right\rangle  \\
\Phi_l  &= -\left\langle \boldsymbol{\nabla}\widetilde{\boldsymbol{u}}:\overline{\mathbf{P}}\right\rangle\\
\mathcal{D}_l &= \eta\left\langle |\overline{\boldsymbol{J}}|^{2}\right\rangle  
\end{align}
where
$\widetilde{\boldsymbol{\tau}}=\widetilde{\boldsymbol{u}}\widetilde{\boldsymbol{u}}-\widetilde{\boldsymbol{u}\boldsymbol{u}}$.
As before we defined the validity test of the coarse-graining (CG) equation, Eq.~(\ref{eqcgen}), as
\begin{equation}
\mathcal{O}_{l}=\partial_t E_l + \Pi_{\text{MHD}l} +  \Pi_{\text{Hall}l} +\Phi_l +\mathcal{D}_l
\label{cgol}
\end{equation}

\begin{figure}[htb]
\centerline{
\includegraphics[width=7.8cm]{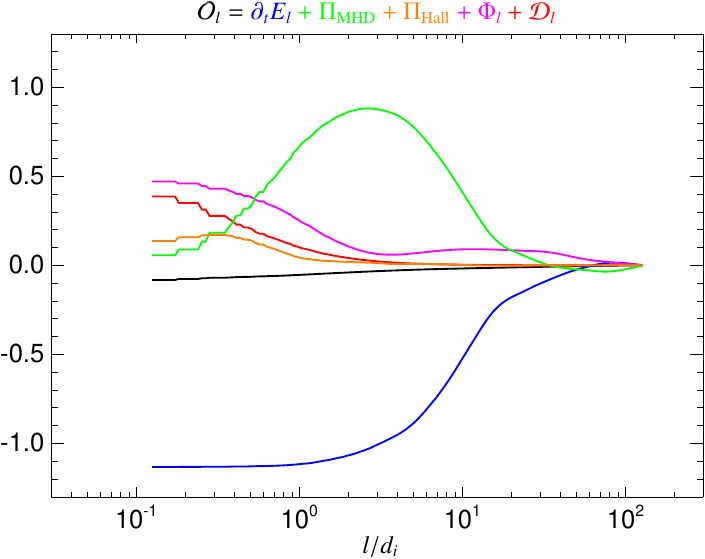}
}
\caption{
 Coarse-graining (CG) analysis of the combined energy:
Validity test of the isotropic
 CG $\mathcal{O}_{l}$, Eq.~(\ref{cgol}),    as a function of $k$
 at $t \Omega_i= 700$(black line) along with
the different contributing terms:
  decay rate $\partial_t  E_l$ (blue),
   MHD cascade rate $\Pi_{\text{MHD}l}$ (green),
   Hall cascade rate $\Pi_{\text{Hall}l}$ (orange),
 resistive dissipation rate $\mathcal{D}_l$ (red), and
 pressure-strain rate $\Phi_l$ (magenta).
All the quantities are given in units of the total (effective) dissipation rate
$Q$.
\label{encg}
}
\end{figure}

Figure~\ref{encg} shows results of the coarse-graining for
the spatial low-pass filter using the normalized boxcar window function.
The CG validity test is relatively well satisfied 
($|\mathcal{O}_{l}|/Q\lesssim 10$ \%); the respective errors of the
KHM, ST, and CG approaches are comparable.
Figure~\ref{encg} repeats the turbulence picture we saw in the KHM and
ST approaches, the energy decays at large
scales, cascades at intermediate scales, and at small scales the
cascade partly continues via the Hall term and partly is dissipated
by the resistive and pressure-strain effects.
For the spatial low-pass filter and the same separation and filter scale $l$
we obtained quantitatively equivalent MHD and Hall cascade rates compared
to the KHM equation (and, consequently, compared to the ST equation
as we saw in the previous subsection):
\begin{equation}
\Pi_{\text{MHD}l} \simeq K_\text{MHD}, \  \
\Pi_{\text{Hall}l} \simeq K_\text{Hall}.
\end{equation}
The other respective terms in the CG and KHM approaches
exhibit a complementarity behaviour, which is similar to that of the ST approach
in subsection \ref{appen1}.
We note that it is possible to use the spectral filtering
in a way analogical to the CG approach \citep{arroal22}.
It is possible to use spectral low-pass filters, which lead to an inverse
dependence of the different quantities on $l$ (i.e. $l\rightarrow -l$). 

\section{Cross helicity}
\label{apphc}

\subsection{Spectral transfer}
\label{apphc1}
For the spectral scale decomposition of the cross helicity we chosen 
\begin{align}
H_{ck}&=\Re\sum_{|\boldsymbol{k}^\prime|\le k}
\left(\widehat{\boldsymbol{u}}^*\cdot \widehat{\boldsymbol{B}_1}\right)
\end{align}
In analogy with the KHM equation (where
the increment representation losses the information about the
background magnetic field $\boldsymbol{B}_0$)
we used the fluctuating magnetic field $\boldsymbol{B}_1=\boldsymbol{B}-\boldsymbol{B}_0$.
For $H_{ck}$ we
 obtained the following dynamic equation
\begin{align}
\partial_t H_{ck}+{S\!}_{Hck}=-\Psi_{Hck}-D_{Hck},
\label{hcsptrdyn}
\end{align}
where ${S\!}_{Hck}$ 
represents  the (joint MHD and Hall) cross helicity transfer rate,
and $\Psi_{Hck}$ and $D_{Hck}$ describe the pressure-strain and resistive dissipation 
 effects, respectively. These terms can be expressed as
\begin{align}
{S\!}_{Hck}&=\Re\sum_{|\boldsymbol{k}^\prime|\le k}\bigg[
\widehat{\boldsymbol{B}_1}^{*}\cdot\widehat{(\boldsymbol{u}\cdot\boldsymbol{\nabla})\boldsymbol{u}} 
-\widehat{\boldsymbol{B}_1}^{*}\cdot\widehat{\frac{\boldsymbol{J}\times\boldsymbol{B}}{\rho}} \nonumber \\
&-\widehat{\boldsymbol{\omega}}^{*}\cdot\widehat{\boldsymbol{u}\times\boldsymbol{B}} 
+\widehat{\boldsymbol{\omega}}^{*}\cdot\widehat{\boldsymbol{j}\times\boldsymbol{B}} \bigg] \\
\Psi_{Hck}&=\Re\sum_{|\boldsymbol{k}^\prime|\le k} \widehat{\boldsymbol{B}_1}^{*}\cdot\widehat{\frac{\boldsymbol{\nabla}\cdot\boldsymbol{\mathrm{P}}}{\rho}} \\
D_{Hck}&=\eta\Re\sum_{|\boldsymbol{k}^\prime|\le k}  |\boldsymbol{k}^\prime|^{2} \widehat{\boldsymbol{u}}^{*}\cdot\widehat{\boldsymbol{B}}
\end{align}
Here again we defined the validity test of the ST cross-helicity equation (\ref{hcsptrdyn}) as
\begin{align}
O_{Hck}=\partial_t H_{ck}+{S\!}_{Hck}+\Psi_{Hck}+D_{Hck}.
\label{okhc}
\end{align}

\begin{figure}[htb]
\centerline{
\includegraphics[width=7.8cm]{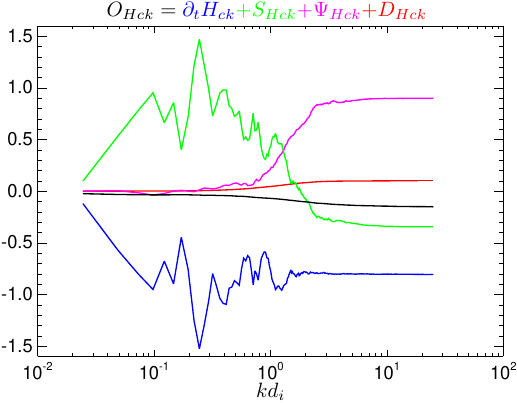}
}
\caption{
ST analysis of the cross helicity:
Validity test of the isotropic
ST $O_{Hck}$, Eq.~(\ref{okhc}),    as a function of $k$
at $t \Omega_i = 700$ (black line)
along with
the different contributing terms:
  decay rate $\partial_t H_{ck}$ (blue),
   cascade rate ${S\!}_{Hck}$ (green),
 resistive dissipation rate $D_{Hck}$ (red), and
 pressure-strain rate $\Psi_{kHc}$ (magenta).
All the quantities are given in units of the total (effective) cross-helicity dissipation rate
$\epsilon_{Hc}$.
\label{hcsptr}
}
\end{figure}

Figure~\ref{hcsptr} shows that the cross helicity ST equation (\ref{hcsptrdyn})
is relatively well satisfied ($|O_{Hck}|/\epsilon_{Hc}\lesssim 15 \%$).
In the Fig.~\ref{hcsptr} we recovered the KHM results,
the cross helicity decays at large
scales, cascades at intermediate scales, and at small scales 
it is dissipated by the resistive and pressure-strain effects, with the latter process
being the dominant one. 
Similarly to the case of the energy ST and KHM equations the 
cross helicity ST and KHM approaches give equivalent cascade rates
\begin{equation}
 {S\!}_{Hck}  \simeq K_\text{Hc}
\end{equation}
for $k l = \sqrt{2}$.
The other terms have again complementary behaviour.

\subsection{Coarse graining}
\label{apphc2}

For the scale-decomposition of the cross helicity
we choose the following filtered quantity
\begin{equation}
H_{cl}=\left\langle \overline{\boldsymbol{B}_1}\cdot\overline{\boldsymbol{u}}\right\rangle.
\end{equation}
For the coarse-grained cross helicity $H_{cl}$
we obtained this dynamic equation
\begin{equation}
\partial_t H_{cl} + \Pi_{Hcl}  = -\Phi_{Hcl} -\mathcal{D}_{Hcl}.
\label{eqcghc}
\end{equation}
where $\Pi_{Hcl}$
represents  the cross helicity transfer rate,
and $\Phi_{Hcl}$ and $\mathcal{D}_{Hcl}$ describe the pressure-strain and resistive dissipation 
 effects, respectively. These terms can be expressed as
\begin{align}
 \Pi_{Hcl} & = \left\langle \overline{\boldsymbol{B}_1}\cdot\overline{\left(\boldsymbol{u}\cdot\boldsymbol{\nabla}\right)\boldsymbol{u}}\right\rangle 
-\left\langle \overline{\boldsymbol{B}_1}\cdot\overline{\frac{\boldsymbol{J}\times\boldsymbol{B}}{\rho}}\right\rangle 
-\left\langle \overline{\boldsymbol{\omega}}\cdot\overline{\boldsymbol{u}\times\boldsymbol{B}}\right\rangle\nonumber \\ 
&+\left\langle \overline{\boldsymbol{\omega}}\cdot\overline{\boldsymbol{j}\times\boldsymbol{B}}\right\rangle \\
\Phi_{Hcl} &= \left\langle \overline{\boldsymbol{B}_1}\cdot\overline{\frac{\boldsymbol{\nabla}\cdot\mathbf{P}}{\rho}}\right\rangle  \\
\mathcal{D}_{Hcl}&=\eta\left\langle \overline{\boldsymbol{\omega}}\cdot\overline{\boldsymbol{J}}\right\rangle  
\end{align}
As before, we defined the validity test of the CG cross-helicity equation (\ref{eqcghc}) as
\begin{equation}
\mathcal{O}_{Hcl}=\partial_t H_{cl} + \Pi_{Hcl}  +\Phi_{Hcl} +\mathcal{D}_{Hcl}
\label{gcohcl}
\end{equation}
\begin{figure}[htb]
\centerline{
\includegraphics[width=7.8cm]{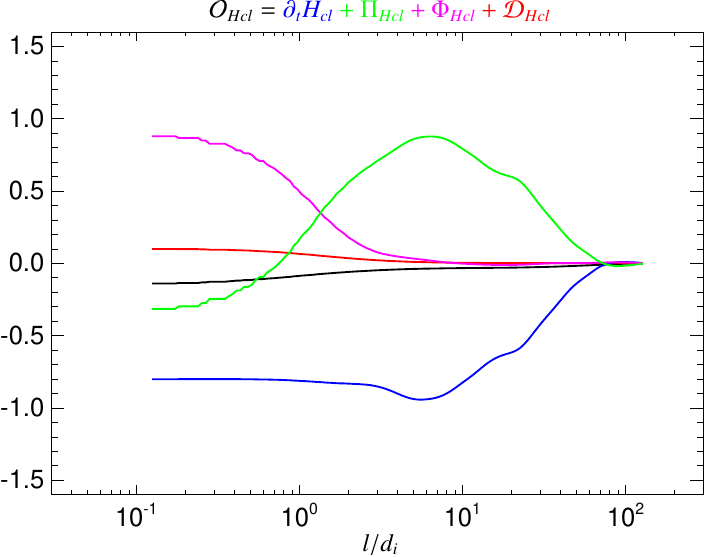}
}
\caption{
 CG analysis of the cross helicity:
Validity test of the isotropic
 CG $\mathcal{O}_{Hcl}$, Eq.~(\ref{gcohcl}),    as a function of $k$
at  $t \Omega_i = 700$ (black line)
along with the different contributing terms:
  decay rate $\partial_t  H_{cl}$ (blue),
   MHD cascade rate $\Pi_{Hcl}$ (green),
 resistive dissipation rate $\mathcal{D}_{Hcl}$ (red), and
 pressure-strain rate $\Phi_{Hcl}$ (magenta).
All the quantities are given in units of the total (effective) cross-helicity dissipation rate
$\epsilon_{Hc}$.
\label{chcg}
}
\end{figure}
Figure~\ref{chcg} shows that the cross helicity CG equation is
relatively well satisfied ($|\mathcal{O}_{Hcl}|/\epsilon_{Hc}\lesssim 16 \%$).
Figure~\ref{chcg} again repeats the same information seen in the corresponding
KHM and ST analyses: the cross helicity decays at large
scales, cascades at intermediate scales, and at small scales
it is dissipated by the resistive and pressure-strain effects.
The KHM and CG approaches give quantitatively similar results,
for the cascade rates
\begin{equation}
\Pi_{Hcl} \simeq K_{Hc}.
\end{equation}
The other respective terms in the CG and KHM approaches
exhibit a complementarity behaviour, which is similar to that of the ST approach
in subsection \ref{apphc1}.

\end{appendix}
\end{document}